\documentclass[conference]{IEEEtran}

\usepackage[dvips]{color}
\usepackage{epsf}
\usepackage{times}
\usepackage{epsfig}
\usepackage{graphicx}
\usepackage{pstricks}
\usepackage{amsmath}
\usepackage{amssymb}
\usepackage{epstopdf}

\usepackage{amsxtra}
\usepackage{here}
\usepackage{rawfonts}
\usepackage{times}
\usepackage{url}
\usepackage{cite}
\usepackage{comment}
\usepackage[utf8]{inputenc}
\usepackage{subcaption}

\newtheorem{definition}{\bf Definition}
\newtheorem{remark}{\bf Remark}
\ifCLASSINFOpdf
\else
\fi

\IEEEoverridecommandlockouts

\begin{document}

\title{Quantum Game Theory for Beam Alignment in Millimeter Wave Device-to-Device Communications \vspace{-1.6cm}}

\author{\IEEEauthorblockN{Qianqian Zhang$^1$, Walid Saad$^1$, Mehdi Bennis$^2$, and M\'erouane Debbah$^{3,4}$}
	
	\IEEEauthorblockA{\small
		$^1$Bradley Department of Electrical and Computer Engineering, Virginia Tech, Blacksburg, VA, USA,
		Emails: \url{{qqz93,walids}@vt.edu}\\
		$^2$Center for Wireless Communications-CWC, University of Oulu, Finland, Email:\url{bennis@ee.oulu.fi}\\
		$^3$Mathematical and Algorithmic Sciences Lab, Huawei France R\&D, Paris, France, Email:\url{merouane.debbah@huawei.com}\\
		$^4$Large Systems and Networks Group (LANEAS), CentraleSupélec, Université Paris-Saclay, 3 rue Joliot-Curie, 91192 Gif-sur-Yvette, France \thanks{This work was supported by the U.S. National Science Foundation under Grants CNS-1526844, CNS-1460316, and CNS-1513697, and by the Office of Naval Research (ONR) under Grant N00014-15-1-2709.}
		\vspace{-0.4cm}		
}
}
\maketitle

\begin{abstract}

In this paper, the problem of optimized beam alignment for wearable device-to-device (D2D) communications over millimeter wave (mmW) frequencies is studied. In particular, a noncooperative game is formulated between wearable communication pairs that engage in D2D communications. In this game, wearable devices acting as transmitters autonomously select the directions of their beams so as to maximize the data rate to their receivers. To solve the game, an algorithm based on best response dynamics is proposed that allows the transmitters to reach a Nash equilibrium in a distributed manner.
To further improve the performance of mmW D2D communications, a novel \emph{quantum game} model is formulated to enable the wearable devices to exploit new quantum directions during their beam alignment so as to further enhance their data rate.
Simulation results show that the proposed game-theoretic approach improves the performance, in terms of data rate, of about $75\%$ compared to a uniform beam alignment. The results also show that the quantum game model can further yield up to $20\%$ improvement in data rates, relative to the classical game approach.
\end{abstract}

\IEEEpeerreviewmaketitle

\vspace{-0.1cm}
\section{Introduction \label{sec1}}

Wearable devices are rapidly becoming an integral part of future communication networks \cite{5G05}. Such devices can provide a variety of services such as health monitoring, entertainment and office helper.
One promising approach for deploying wearables in future wireless networks is via the use of D2D communications \cite{D2DmmW01}. In particular, D2D communication over mmW frequencies ranging from $30$~GHz to $300$~GHz can provide the suitable platform for wearable networks.

Operating D2D over mmW for wearable communications faces a number of challenges, such as the high attenuation loss and low diffraction efficiency of mmW \cite{D2DmmW03}.
As shown in \cite{5G05}, compared with conventional sub-6 GHz signals, mmW at the $30$~GHz experiences an additional 20 dB attenuation loss regardless of the transmission distance. Further, the low diffraction efficiency caused by the short wavelength of mmW reduces its ability to penetrate common obstacles, such as the human body \cite{D2DmmW01}.
To adapt to such a harsh propagation environment, directional antenna arrays can be used for mmW D2D devices. Beamforming performed by antenna arrays can provide a high gain to compensate for serious attenuation, as well as improve the channel condition by searching for the line-of-sight link.
However, one challenge for beamforming is known as \emph{deafness} in which the antenna beams of the transmitter and receiver are not aligned \cite{beams}. We will study this problem in our work.

Most of previous works \cite{D2DmmW01,D2DmmW03} focus on the performance of D2D mmW communication with the assumption of perfect beam alignment.
The authors in \cite{xin} incorporate the impact of beam association into system performance using stochastic geometry. However, this work does not provide any practical approach for the users to dynamically adapt their beam alignment.
Some recent works \cite{beamAlign01,beamAlign02} study the physical realization of beam alignment, but no considerations from a system level has been presented. In this work, we will come up with a distributed approach for each wearable device to implement the beam alignment by detecting the current network state and interacting between multiply transmitter-receiver pairs.

Beyond intelligently optimizing the beam direction, one emerging technology for improving beamforming in communications is via the use of \emph{quantum models} \cite{xin}. For instance, the authors in \cite{quanAnte} have developed a quantum antenna model that allows quantum beam alignment on wearable devices.
For quantum antenna arrays, the degrees of freedom of the beam alignment space can be largely expanded.
Some additional methods for antenna deployment, which exist \emph{only in the quantum domain}, can be leveraged.
Clearly, quantum models have a promising potential to significantly improve the overall data rates of wearable communications over mmW.

The main contribution of this paper is to introduce a novel framework for optimizing beam alignment for mmW D2D communication. To this end, we formulate the problem as a noncooperative game in which the players are the wearable transmitters, who seek suitable D2D beam directions to optimize their own data rates.
This distributed beamform alignment model is further enhanced by introducing a novel approach that leverages tools from quantum game theory (QGT) \cite{QGT01}, which is shown to further improve the performance of the system.
To the best of our knowledge, this paper \emph{is the first to develop a quantum game model} for addressing the beam alignment problem in mmW D2D networks.
To solve both the conventional and quantum games, we propose a distributed algorithm, based on best response dynamics, which enables the transmitters to find their Nash equilibrium strategies. Simulation results show that the proposed game-theoretic approach significantly improves the data rate by up to $75\%$ compared to a uniform beam alignment strategy.
Further, the QGT approach can yield up to $20\%$ of capacity improvement, relative to the standard game.

The rest of this paper is organized as follows. Section~\ref{sec2} presents the system model. The problem formulation and solution of the beam alignment are proposed in Section~\ref{sec3}. Section~\ref{sec4} presents the quantum game formulation and its feasible solution. Simulation results are presented in Section~\ref{sec5}. Finally, conclusions are drawn in Section~\ref{sec6}.

\section{System Model \label{sec2}}

Consider a two-dimensional network with $N$ users, each of which is equipped with a transmitter-receiver pair of D2D communication devices. Let $\mathcal X$ be the set of $N$ transmitters and $\mathcal Y$ be the set of $N$ receivers.

\subsection{Blockage and Channel Model}
To model the blockage effect of the human body, a circle is used to represent the blockage area. The wearable device is located around the human body at a distance
from the edge of the blockage circle.
If the connecting line between a transmitter and a receiver does not pass through any blockage area, the communication link is said to be line-of-sight (LOS); otherwise, the link is non-line-of-light (NLOS).
Accounting for path loss and shadow fading, the attenuation equation has the following form \cite{Omid}:
\begin{align}
h[dB]=a+20\log_{10} (f)+10n\log_{10} (d)+z_\sigma,
\label{equation1}
\end{align}
where $a$ is the path loss parameter, $f$ denotes the carrier frequency in GHz, $n$ is the path loss exponent, $d$ represents the distance between transmitter and receiver in meters and $z_\sigma$ is a shadow fading term. The values of $a$ and $n$ vary according to the state of link $ij$. The typical parameter values for both LOS and NLOS cases are available in IEEE 802.11ad \cite{standard}.

\subsection{Antenna Model}
\begin{figure}[!t]
\begin{center}
\vspace{-0.5cm}
\includegraphics[width=6cm]{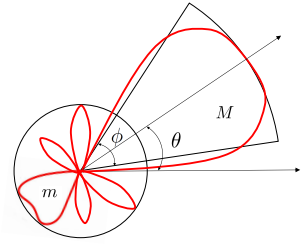}
\vspace{-0.1cm}
\caption{\label{at} \small Directional beamforming is performed only at transmitters. The antenna model is characterized by four parameters: the boresight angle $\theta$, the beamwidth $\phi$, the mainlobe gain $M$ and the sidelobe gain $m$.
}
\end{center}\vspace{-1cm}
\end{figure}

For tractability, only transmitters are deployed with directional antennas, while antennas on receivers are omnidirectional.
As illustrated in Fig.~\ref{at}, the directional antenna is simplified to a two-dimensional model with four parameters: $\phi$, $\theta$, $M$, $m$, where $\phi$ is the half-power beamwidth, $\theta$ is the boresight direction, $M$ and $m$ are the antenna gains of the mainlobe and the sidelobe, respectively. This model is commonly used in D2D mmW literature such as \cite{D2DmmW01} and \cite{D2DmmW03}.
If a receiver is located within the mainlobe of a transmitter, the antenna gain of the link is a constant of $M$; otherwise the gain is $m$.
Let $\varphi_{ij}$ denote the phase of vector from the position of transmitter $i$ to receiver $j$.
Then, the antenna gain of the link between transmitter $i$ and receiver $j$ can be expressed as \cite{D2DmmW03}:
\vspace{-0.1cm}
\begin{eqnarray}g_{ij}=
\begin{cases}
M, &|\varphi_{ij} -\theta_i |\leq{\frac{\phi_i}{2}}, \cr m, & |\varphi_{ij} - \theta_i|>{\frac{\phi_i}{2}}, \end{cases}
\end{eqnarray}

\subsection{Interference Model}
Given the assumptions above, the co-channel signal to interference plus noise ratio (SINR) for a typical communication link of transmitter $i$ and receiver $j$ is \cite{D2DmmW01}:
\begin{align}
\Gamma_{ij} = \dfrac{g_{ij} h_{ij} P_{ij}}{\sigma^{2} + \sum_{k=1, k \neq j}^{N} g_{kj} h_{kj} P_{kj}},
\label{equation13}
\end{align}\vspace{-0.15cm}\\
where $h_{ij}$, given by $(\ref{equation1})$, corresponds to the path loss from transmitter $i$ to receiver $j$ in linear scale, $\sigma^{2}$ represents a constant thermal noise power, and $P_{ij}$ is the transmit power.
Hereinafter, we assume a fixed transmit power, thus let $P_{ij} = P,\ \forall i \in \mathcal{X}, j\in\mathcal{Y}$. Since the channel attenuation loss is determined by the transmitter-receiver location and the environment, in order to improve the transmission SINR given in (\ref{equation13}),
each transmitter can adjust the antenna gains by exploring different beam directions.

\section{Game Formulation and Solution \label{sec3}}

\subsection{Problem formulation \label{ssec31}}
We study this beam alignment problem using the framework of game theory \cite{GT01}. A static noncooperative game is formulated in strategic form as $\mathcal G = (\mathcal X, \{\Theta_i\}_{i \in \mathcal X},\{U_i\}_{i \in \mathcal X})$,
where (i) $\mathcal X$, the set of \emph{players}, denotes the transmitters,
(ii) $\{\Theta_i\}$, the \emph{strategies} space for each player $i \in \mathcal X$, corresponds to the set of the beamforming directions $\theta_i \in \Theta_i = [0,\phi_i, 2\phi_i, \cdots, 2\pi-\phi_i]$.
For each transmitter $i \in \mathcal{X}$,
the value of beamwidth $\phi_i$ must exactly divide $2\pi$ so that every direction within $[0,2\pi)$ can be covered by the mainlobe when choosing one specific strategy, and (iii) the \emph{utility function} $U_i$ is defined as the capacity of the communication link from transmitter $i$ to its receiver. The utility for transmitter $i$ when choosing strategy $\theta_i$ is given by:
\begin{align} \vspace{-0.05cm}
U_i(\theta_i,\boldsymbol{\theta}_{-i}) = \log_2\left(1 + \dfrac{g_{ii}(\theta_i) h_{ii} P}{\sigma^{2} + \sum_{j\in \mathcal X,j\neq i} g_{ji}(\boldsymbol{\theta}_{-i}) h_{ji} P}\right),
\label{equation2}
\end{align}
where $\boldsymbol{\theta}_{-i} := [\theta_1, \cdots, \theta_{i-1}, \theta_{i+1}, \cdots, \theta_N]$ is the vector of the strategies of the other $N-1$ players except $i$. $\boldsymbol \theta$ denotes the $ N\times1$ vector of strategies for all players in $\mathcal X$. Here, the objective for each transmitter $i \in \mathcal X$ is to choose a beam direction $\theta_i$ so as to optimize its transmission capacity given in $(\ref{equation2})$.

\subsection{Game Solution \label{ssec32}}
One popular solution for a noncooperative game is the so-called \emph{Nash equilibrium} (NE) defined next.
\begin{definition}
A \emph{Nash equilibrium} of a static noncooperative game $\mathcal G = (\mathcal X, \{\Theta_i\}_{i \in \mathcal X},\{U_i\}_{i \in \mathcal X})$ defined in Section~\ref{ssec31}, with utility given in $(\ref{equation2})$ is a vector of action $\boldsymbol{\theta}^* \in \Theta$ such that $\forall i \in \mathcal X$, the following holds:
\begin{align}
U_i(\theta_i^*, \boldsymbol{\theta}_{-i}^*) \geq U_i(\theta_i, \boldsymbol{\theta}_{-i}^*) , \ \ \forall \theta_i\in{\Theta}_i.
\label{equation3}
\end{align}
\end{definition}

The NE of the beam alignment problem represents a stable state in which no transmitter can improve its data capacity by unilaterally deviating from the current choice of a beam direction, given that the boresight directions of the other transmitters remain fixed.


In the proposed game, once the strategy profile $\boldsymbol \theta_{-i}$ for the other transmitters is determined, transmitter $i$ can evaluate its potential payoff of each strategy $\theta_i \in \Theta_i$ based on the utility function $(\ref{equation2})$.
Therefore, to maximize the data rate, each transmitter can choose an optimal beam direction by calculating its \emph{best response} (BR).
The best response  $\emph{b}_i(\boldsymbol{\theta_{-i}})$ of player $i$ to the strategy profile $\boldsymbol{\theta}_{-i}$ is a set of strategies such that $b_i(\boldsymbol{\theta}_{-i})\!=\!\{\theta_i \in {\Theta}_i|U_i(\theta_i,\boldsymbol{\theta}_{-i}) \ge U_i(\theta_i^\prime,\boldsymbol{\theta}_{-i}),\ \forall \theta_i^\prime \in {\Theta}_i\}$, which means, the achievable data rate through $b_i(\boldsymbol{\theta}_{-i})$ is at least as good as the performance of any other strategies in $\Theta_i$.

Note that, the best response $\emph{b}_i(\boldsymbol{\theta_{-i}})$ for each transmitter $i\in \mathcal{X}$ is always the beam direction $\theta_i$, which guarantees $g_{ii}(\theta_i)=M$.  According to (\ref{equation2}), the strategy choice $\theta_i$ only affects the value of antenna gain $g_{ii}(\theta_i)$ in the utility function. In order to maximize the data capacity, each transmitter can search for the optimal beam direction, which enables its receiver to be included into the mainlobe of its transmission antenna, to get the maximal antenna gain. 
Therefore, the NE of the proposed game $\mathcal{G}$ always exists and is defined as $\boldsymbol{\theta}^* \in \boldsymbol{\Theta}$, where $|\varphi_{ii} -{\theta_i}^* |\leq{\frac{\phi_i}{2}}$, $\forall i \in \mathcal{X}$. Indeed, the set of possible values for $\boldsymbol{\theta}^*$ is also the set of the dominant strategies for all transmitters.

\section{quantum game model and solution \label{sec4}}

Using the beam alignment approach proposed in Table~\ref{tab:algo}, each transmitter can deploy its beam direction strategically to optimize the transmission capacity.
However, the data rate resulting from the NE is not always efficient for the system.
To overcome this inefficiency, one can explore \emph{quantum strategies} to reach a more favorable game solution compared with the pure NE \cite{QGT01}.
A quantum approach for beam alignment can enable an antenna to exploit \emph{directions in the quantum space}, which can potentially improve the overall pairing between transmit and receive antennas. How to integrate such quantum directions into the beam alignment game is the focus of the following work. Here, due to space limitations, not all the basic quantum calculations \cite{quanIntro2} can be covered. Thus, we restrict our discussion to the key notions needed in our models.

\vspace{-0.05cm}
\subsection{Quantum Game Model \label{ssec41}}
\vspace{-0.02cm}
To perform the quantum analysis, we consider a model analogous to Section III, however, the antennas are assumed to perform quantum alignment in a quantum state \cite{quanBook}.
To clearly showcase the benefits of the quantum game model, hereinafter, we restrict our attention to a case of two players and three strategies.

In a conventional game, since each transmitter $i \in \mathcal X$ has three strategy choices, the beamwidth of the directional antenna for each player is $ \phi_i=2\pi/3$, thus the strategy space is $\boldsymbol{\theta}_i = [0, 2\pi/3, 4\pi/3]$.
By selecting one strategy $\theta_i \in \boldsymbol{\theta}_i$, transmitter $i$ would set its antenna boresight to angle $\theta_i$.
And we refer to this deployment as preparing the \emph{game state} $\mu_i$ of transmitter $i$ into $\mu_i = \theta_i$.
In the quantum game, the game state of each player $i$ is described by a Dirac notation $|\mu_i\rangle$ \cite{quanIntro2}, which is mathematically denoted as a vector. For example, by setting the beam direction to be $0$,
the transmitter $i$ in the quantum case would prepare its game state into $|\mu_i\rangle = [1,0,0]^\mathrm{T}$. The choice of beam direction $2\pi/3$ or $4\pi/3$ would change the game state into $|\mu_i\rangle = [0,1,0]^\mathrm{T}$ or $[0,0,1]^\mathrm{T}$, respectively.
Let $\mathcal E = \{[1,0,0]^\mathrm{T}, [0,1,0]^\mathrm{T}, [0,0,1]^\mathrm{T}\}$ be the domain of basic quantum game state for each transmitter.

Another remarkable trait for a quantum state is uncertainty \cite{quanBook}, which stems from the quantum mechanics principle. In most cases, multiple classical and certain states can \emph{superpose} \cite{quanIntro2} with each other and coexist in one quantum state $|\mu_i\rangle$.
When measured by a Stern Gerlach detector \cite{quanBook}, the quantum beam deployment would ``collapse'' into one of the classical forms.
We mathematically represent the superposed quantum state by a linear combination of basic classical states. Then, the probability of collapsing to a classical state will be equal to the conjugate product of its coefficient in the linear combination \cite{quanIntro2}. For example, the quantum game state $|\mu_i\rangle = \alpha\cdot[1,0,0]^T + \beta\cdot [0,1,0]^T$, where $|\alpha|^2+|\beta|^2=1$, may either collapse to the classical game state $0$ with the probability of $|\alpha|^2$, or $2\pi/3$ with the probability of $|\beta|^2$.
Note that, a basic quantum state $|\mu_i\rangle \in \mathcal E$ always reduces into its corresponding classical form every time being detected,
and thus, it is deterministic and is physically equivalent to that classical state.
The inclusion of these certain states enables the quantum model to express classical game faithfully.
Furthermore, the notion of the quantum game state for the entire network is denoted as $|\mu_1 \mu_2 \rangle = |\mu_1 \rangle \otimes |\mu_2 \rangle$, which is the tensor product of every transmitter's game state \cite{quanBook}.

Consequently, a quantum game is formulated in the strategic form as $\mathcal Q = (\mathcal X, \{\hat\Theta_i\}_{i \in \mathcal X},\{\hat U_i\}_{i \in \mathcal X})$, where $\mathcal X$, the set of players, are two transmitters;
$\{\hat\Theta_i\}$ is the strategy set for each transmitter $i\in\mathcal X$, which, in the quantum domain, is described by matrices.
Here, a  $3 \times 3$ unitary matrix with 3 parameters $(\alpha, \beta, \lambda)$ is used to present the quantum strategy as:
\begin{align} \vspace{-0.02cm}
\hat{\Theta}_i(\alpha,\beta,\lambda)=\left(\begin{array}{cccc}
\text{cos}\alpha & -\text{sin}\alpha & 0\\
\text{sin}\alpha \cdot \text{cos}\beta & \text{cos}\alpha \cdot \text{cos}\beta & -\text{sin}\beta \cdot e^{i\lambda}\\
\text{sin}\alpha \cdot \text{sin}\beta & \text{cos}\alpha \cdot \text{sin}\beta & \text{cos}\beta \cdot e^{i\lambda}\end{array}\right)
,\vspace{-0.5cm}
\label{equation6}
\end{align}
where $\alpha, \beta, \lambda \in [0,\pi/2]$ are the superposition parameters. In this work, we restrict our analysis on a subset $\mathcal{S} = \mathcal{S}_1 \cup \mathcal{S}_2 \cup \mathcal{S}_3 = \{0\le\alpha\le\pi/2, \beta = 0, \lambda = 0\} \cup \{\alpha = \pi/2, 0\le\beta\le\pi/2, \lambda = 0\} \cup \{\alpha = 0, \beta = 0, 0\le\lambda\le\pi/2\}$, where the quantum strategy $\hat{\Theta}_i (\alpha, \beta, \lambda)$ has a specific physical meaning.
From (\ref{equation6}),
we can see that the strategy set of each transmitter $i\in\mathcal X$ is extended from a vector of three discrete elements into a $3 \times 3$ matrix spanning in three dimensions.

Next, we demonstrate the physical layout of antennas by applying the quantum strategies
and the relationship of strategy spaces between the classical $\boldsymbol \theta$ and the quantum $\hat{\boldsymbol{\Theta}}$.
First, the quantum strategy, which corresponds to $\theta_i = 0$ of the classical domain, is defined as,
$\Delta = \hat{\Theta}_i(0,0,0)$, which is the $3 \times 3$ identity matrix.
If transmitter $i$ applies quantum strategy $\Delta$, its antenna boresight direction is physically deployed at $0$. And the quantum game state for transmitter $i$ will map into $|\mu_i\rangle = [1,0,0]^\mathrm{T}$.
If transmitter $i$ applies the quantum strategy $\Upsilon = \hat{\Theta}_i(\pi/2,0,0)$ or $\Omega = \hat{\Theta}_i(\pi/2,\pi/2,0)$, the beam direction becomes $2\pi/3$ or $4\pi/3$, and the game state is prepared as $|\mu_i\rangle = [0,1,0]^\mathrm{T}$ or $|\mu_i\rangle = [0,0,1]^\mathrm{T}$, respectively.
Note that, the quantum game state $|\mu_i\rangle$ for each transmitter is determined by the quantum strategy $\hat\Theta_i$, while the quantum strategy not only decides the game state of its transmitter, but also influences the game state of the entire networks.

\begin{figure}[!t]
	\begin{center}
		\vspace{-1.2cm}
		\includegraphics[width=8cm]{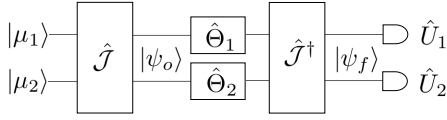}
		\vspace{-1.9cm}
		\caption{\label{qgt}\small The diagram shows the system of a two-player quantum game, which starts with the initial system states $|\mu_1 \mu_2 \rangle$, and ends with the expected utilities $\hat U_1$ and $\hat U_1$ of two players, respectively. }
	\end{center}\vspace{-0.8cm}
\end{figure}

Beyond $\Delta$, $\Upsilon$, and $\Omega$, which have specific counterparts in the conventional game, the other strategies defined on $\mathcal{S}$ do not explicitly have a traditional physical meaning.
The mathematical analysis reveals that the strategies defined on $\mathcal{S}_1$ is $\hat {\Theta}_i(\alpha,0,0) = $ cos$\alpha \cdot \Delta + $ sin$\alpha \cdot \Upsilon$, which is the superposition of $\Delta$ and $\Upsilon$.
By applying the superposed quantum strategies,
transmitter $i$ places its beamforming direction in a quantum manner, where the antenna may either fall into the strategy $\Delta$ with probability cos$^2{\alpha}$ or strategy $\Upsilon$ with probability sin$^2{\alpha}$ when being detected.
Similarly, the strategies on $\mathcal{S}_2$, which will also put the antenna in a quantum state, are formed by the superposition of $\Upsilon$ and $\Omega$.

Next, we introduce the strategies defined on $\mathcal{S}_3 = \{\alpha = 0, \beta = 0, 0\le\lambda\le\pi/2\}$, which correspond to more special physical deployments of beam directions, compared with the quantum strategies defined before.
The strategy $\Lambda = \hat{\Theta}_i(0,0,\pi/2)$ describes a pure and certain quantum state, which has no counterpart in the classical domain and will never collapse into any basic classical state by detection. By introducing the quantum strategy $\Lambda$, a new dimension is created for the strategy space $\boldsymbol{\hat{\Theta}}$ beyond the two dimensional plane, and the beamforming direction will be placed in the newly created space where two transmitters can share certain information and realize a cooperative deployment to achieve better utilities.
Further, the quantum strategies defined on  $\mathcal{S}_3=\{\alpha = 0, \beta = 0, 0\le\lambda\le\pi/2, \}$ are generated from the superposition of $\Delta$ and $\Lambda$ with the probability cos$^2{\lambda}$ that collapses into $\Delta$ and the probability sin$^2{\lambda}$ that falls into $\Lambda$.

\begin{remark} The implementation of quantum strategies is seemingly analogous to classical \emph{mixed strategies}. However, they are intrinsically different.
QGT requires the players to be capable of presenting a quantum state,
which is uncertain because of the superposition, but can become a deterministic state automatically after external detection (i.e., measurement), while mixed strategies select the beam direction according to a probability distribution.
The only similarity is the fact that quantum mechanics is based on probability theory. Moreover, the quantum strategy $\Lambda$ has no counterpart in classical mixed strategies, which further bolsters the novelty of QGT. 
\end{remark}

%
To implement the quantum game, a unitary operator $\mathcal{\hat J}$ is introduced, which combines the space-separated transmitters to form an integrated system and realizes an \emph{entangled} game state.
Quantum entanglement is a physical phenomenon that occurs for a group of particles when the performance of one may have an immediate impact on the behavior of the other particles.
Then the choices for each entangled particle are highly related, so that their behaviors will influence jointly to the entangled system and  cannot be described independently or sequentially.
To ensure the original classical game $\mathcal G$ can be faithfully represented in the quantum framework, \cite{QGT01} has shown the unitary operator $\hat{\mathcal J}$ should be defined as $\mathcal {\hat{J}} = \textrm{exp}\{i\gamma \hat{D}\otimes \hat{D}/2\}$, where $\gamma \in [0,\frac{\pi}{2}]$ is a measure for the game's entanglement and  $\hat{D}$ is a unitary matrix given as:\vspace{-0.1cm}
\begin{align}
\hat D = \left(\begin{array}{cccc}
0 & 0 & 1\\0 & 1 & 0\\1 & 0 & 0
\end{array}\right).
\end{align}

For $\gamma = 0$, there is no entanglement between antenna beamforming directions of two transmitters, therefore their strategies can linearly and independently work on the system game state; while $\gamma = \pi/2$ is the maximal entanglement case, where the strategy taken by one transmitter affect the antenna deployment of the other transmitter largely. This ``action at a distance'' can enable the cooperative disposition of beam directions among transmitters even in a noncooperative case, and, consequently improve the utility in the QGT as opposed to a conventional game.

As shown in Fig.~\ref{qgt}, the quantum game starts with an initial network state of $|\mu_1\mu_2\rangle$. Then, it becomes $|\psi_o \rangle = \mathcal{\hat J} \cdot |\mu_1\mu_2\rangle $ after $\hat{\mathcal J}$ realizes the entanglement between transmitters.
Then two transmitters execute their strategies $\hat\Theta_1$ and $\hat\Theta_2$, and the game state of the wearable network changes to $(\hat\Theta_1 \otimes \hat\Theta_2) \cdot \mathcal{\hat J} \cdot |\mu_1\mu_2\rangle$. After that, a measurement operation, including an inverse gate $\mathcal {\hat{J}}^ \dagger$ and a pair of detectors, is implemented to determine the utilities.
The final game state before the detection is expressed as $|\psi_f \rangle =\mathcal{\hat{J}}^{\dagger} \cdot (\hat{\Theta}_1 \otimes \hat{\Theta}_2) \cdot \mathcal{\hat{J}} \cdot |\mu_1\mu_2\rangle$, where $|\psi_f\rangle$ is a superposition of all basic states $|{{\mu_1}^f} {{\mu_2}^f} \rangle$, $|{{\mu_1}^f}\rangle$, $|{{\mu_2}^f} \rangle \in \mathcal E$.
Then, by detection, the final quantum game state $|\psi_f \rangle$ collapses into one of the basic status $|{{\mu_1}^f} {{\mu_2}^f} \rangle$, with the probability of $|\langle {{\mu_1}^f} {{\mu_2}^f} | \psi_f \rangle|^2$, where $|\langle {{\mu_1}^f} {{\mu_2}^f} | \psi_f \rangle|^2$ is the inner product of the vectors $|\psi_f \rangle$ and $|{{\mu_1}^f} {{\mu_2}^f} \rangle$.
Since each element of $\mathcal E$ corresponds to a classical strategy $\theta_i$,
we denote ${R_{{\theta_i{\boldsymbol{\theta}_{-i}}}}}(\hat{\Theta}_1,\hat{\Theta}_2) = |\langle {{\mu_1}^f} {{\mu_2}^f} | \psi_f \rangle|^2$ to represent the probability that the strategy profile of two transmitters collapses into $(\theta_i,\boldsymbol{\theta}_{-i})$.
Consequently, assuming a fixed strategy profile $\boldsymbol{\hat{\Theta}}_{-i}$ of other transmitters, the expected utility of transmitter $i$ when taking strategy $\hat{\Theta}_i$ is expressed as:
\vspace{-0.05cm}
\begin{align}
\hat U_i(\hat{\Theta}_i,\boldsymbol{\hat{\Theta}}_{-i}) = \sum_{\theta_i,\boldsymbol{\theta}_{-i}\in [0, \frac{2\pi}{3}, \frac{4\pi}{3}]} U_i(\theta_i,\boldsymbol{\theta}_{-i})\cdot
{R_{{\theta_i{\boldsymbol{\theta}_{-i}}}}}(\hat{\Theta}_i,\boldsymbol{\hat{\Theta}}_{-i}),\vspace{-0.1cm}
\vspace{-0.5cm}
\label{equation9}\vspace{-1cm}
\end{align}
where $\hat U_i (\hat{\Theta}_i, \boldsymbol{\hat{\Theta}}_{-i})$ and $U_i(\theta_i,\boldsymbol{\theta}_{-i})$ are the utilities of transmitter $i$ in the quantum game and the original game respectively, $\Theta_i$ and $\theta_i$ are the strategy of transmitter $i$ in the quantum and non-quantum domain. The objective for each transmitter $i\in\mathcal X$ in the quantum case is to choose an optimal quantum beamforming strategy $\hat{\Theta}_i$ to maximize the utility $\hat U_i(\hat{\Theta}_i,\boldsymbol{\hat{\Theta}}_{-i})$ given in (\ref{equation9}), which presents the expected data rate that transmitter $i$ can achieve by applying strategy $\hat{\Theta}_i$.

\subsection{Solution of the Quantum Game}
To find the NE for the quantum game $\mathcal Q$, we develop an approach based on best response. 
Initially, each transmitter $i \in \mathcal{X}$ chooses an initial beam alignment strategy randomly.
Then given the strategy profile for the other devices fixed, each transmitter takes turn to deploy its beamforming to the optimal direction by adopting its BR strategy.
The iterative process continues until no players can achieve higher capacities by changing its strategy.
Then, this algorithm converges and the current strategy profile $\boldsymbol{\hat{\Theta}}$ is one NE solution.
Note that, since the measurement can cause the collapse of final quantum game state of the network into an unpredicted classical state, every time a new BR iteration starts, the game state $|\mu_i\rangle$ should be reset to the original state, instead of keeping it as the last collapsing result.
Then, the unitary $\mathcal{\hat J}$ is implemented, and transmitters could evaluate every strategy choice according to the expected utility given in (\ref{equation9}) by applying $\hat{J}^\dagger$ to detect the current game state, and then decide its BR strategy $b_i(\boldsymbol{\hat\Theta}_{-i})$. The QGT BR dynamics continues until an NE is found.

However because of the expansion of the strategy set and the entanglement of strategies between transmitters, it is difficult to ascertain the existence of NE in the quantum domain. Also in the quantum game, BR dynamics are not guaranteed converge, since the players may cycle between similar strategy patterns. 
To deal with such a cycling behavior, a timer $T\in \mathbb{N^+}$ is introduced.
The value of $T$ is properly selected so that the BR dynamics process can reach a NE within $T$ iterations, if such an NE exists.
Further, to deal with the possible multiplicity of NEs, the BR dynamics process will be executed for $L\in \mathbb{N^+}$ times with different initial strategy profile.
The timer begins from $T$ and will be reduced by one each time the BR algorithm is applied by any transmitter in $\mathcal X$.
Whenever an NE is found or the timer reaches $0$, the ongoing process stops. Then, the current strategy and the utility for each transmitter $i\in\mathcal X$ would be recorded into the $L \times N$ matrices $\boldsymbol{\hat{\Theta}}^{f}_i$ and $\boldsymbol{\hat{U}}^{f}_i$, respectively. Subsequently, a new BR process starts with a different initial strategy profile. 
After the BR dynamic process has been executed for $L$ times, a procedure, which is summarized in Table.~\ref{tab:algo}, is executed to choose a final beamforming direction profile $\boldsymbol{\hat{\Theta}}^+$.

\begin{table}[!t]
	\centering
	\caption{\vspace*{-0.4em}\small Best response dynamics for the quantum game\label{table4}}
	\begin{tabular}{p{8.5cm}}
		\hline
		\textbf{repeat for $L$ times}\\
		\hspace*{0.5em}Start the game timer $T$\\
		\hspace*{0.5em}Each transmitter $i \in \mathcal{X}$ selects an initial strategy from $\hat{\Theta}_i$ randomly\vspace*{.1em}\\
		\hspace*{0.75em}\textbf{repeat, sequentially}\vspace*{.2em}\\
		\hspace*{1.5em}Transmitter $i \in \mathcal{X}$ applies BR strategy $b_i(\boldsymbol{\hat{\Theta}}_{-i})$;\vspace*{.1em}\\
		\hspace*{1.5em}Timer is subtracted by one;\vspace*{.1em}\\
		\hspace*{0.75em}\textbf{until} convergence to an NE strategy vector $\boldsymbol{\hat{\Theta}}^*$ \textbf{or} timer expires\vspace*{.2em}\\
		\textbf{end}\vspace*{.2em}\\
		Each transmitter $i\in\mathcal X$ broadcasts the utility vector $\boldsymbol{\hat{\Theta}}^{f}_{i}$ sequentially \vspace*{.2em}\\
		\textbf{If only one NE $\boldsymbol{\hat{\Theta}}^*$ exists}\vspace*{.1em}\\
		\hspace*{0.5em} The solution $\boldsymbol{\hat{\Theta}}^+$ equals to $\boldsymbol{\hat{\Theta}}^*$;\vspace*{.1em}\\
		\textbf{Else if multiple NEs exist}\vspace*{.1em}\\
		\hspace*{0.5em}Take the NE 
		that yields the best average capacity as the solution $\boldsymbol{\hat{\Theta}}^+$;\vspace*{.1em}\\
		\textbf{Else}\vspace*{.1em}\\
		\hspace*{0.5em} Enforcement: Take the strategy profile that enables the best average data\\
		\hspace*{0.5em} rate over the network as the solution $\boldsymbol{\hat{\Theta}}^+$.\vspace*{.1em}\\
		\hspace*{0.4em}\textbf{End}\\
		\hline
	\end{tabular}\label{tab:algo}\vspace{-0.5cm}
\end{table}

\begin{figure*}[!t]
\begin{center}
\begin{subfigure}{.315\textwidth}
\centering
\includegraphics[width=6.33cm]{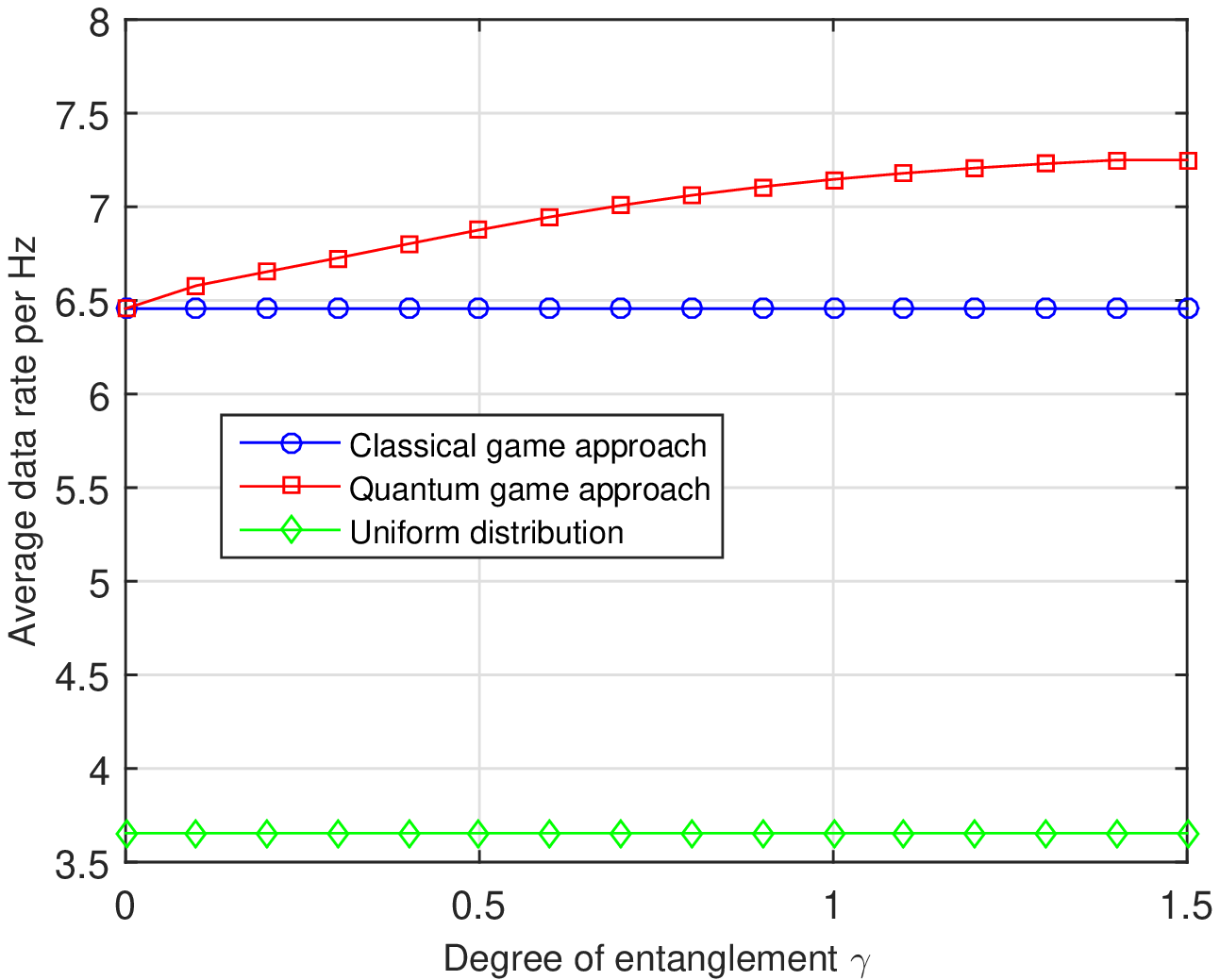}\vspace{-0.2cm}
\caption{\label{entangle}}
\end{subfigure}
\begin{subfigure}{.315\textwidth}
\centering
\includegraphics[width=6.34cm]{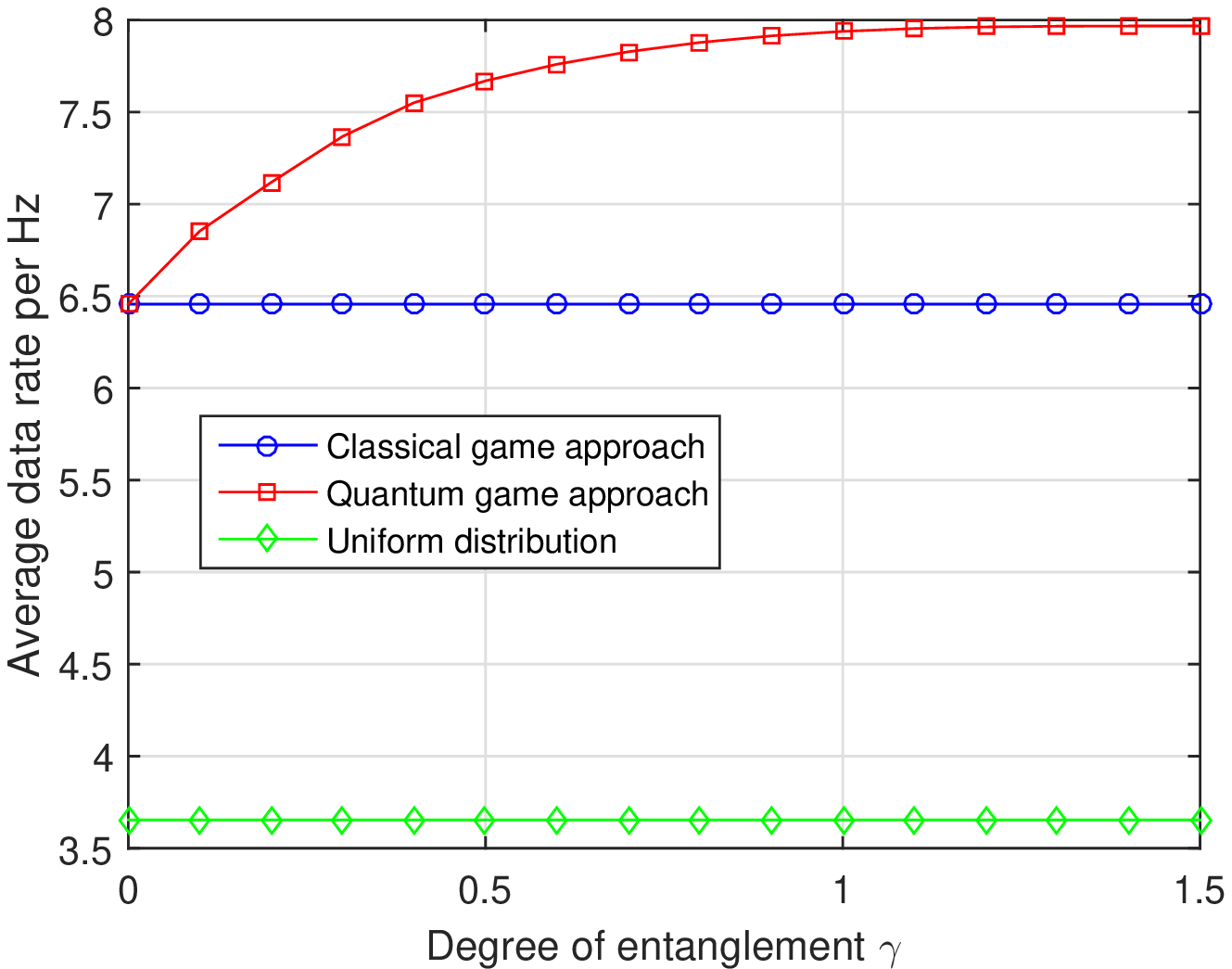}\vspace{-0.2cm}
\caption{\label{only} }
\end{subfigure}
\begin{subfigure}{.31\textwidth}
\centering
\includegraphics[width=6.32cm]{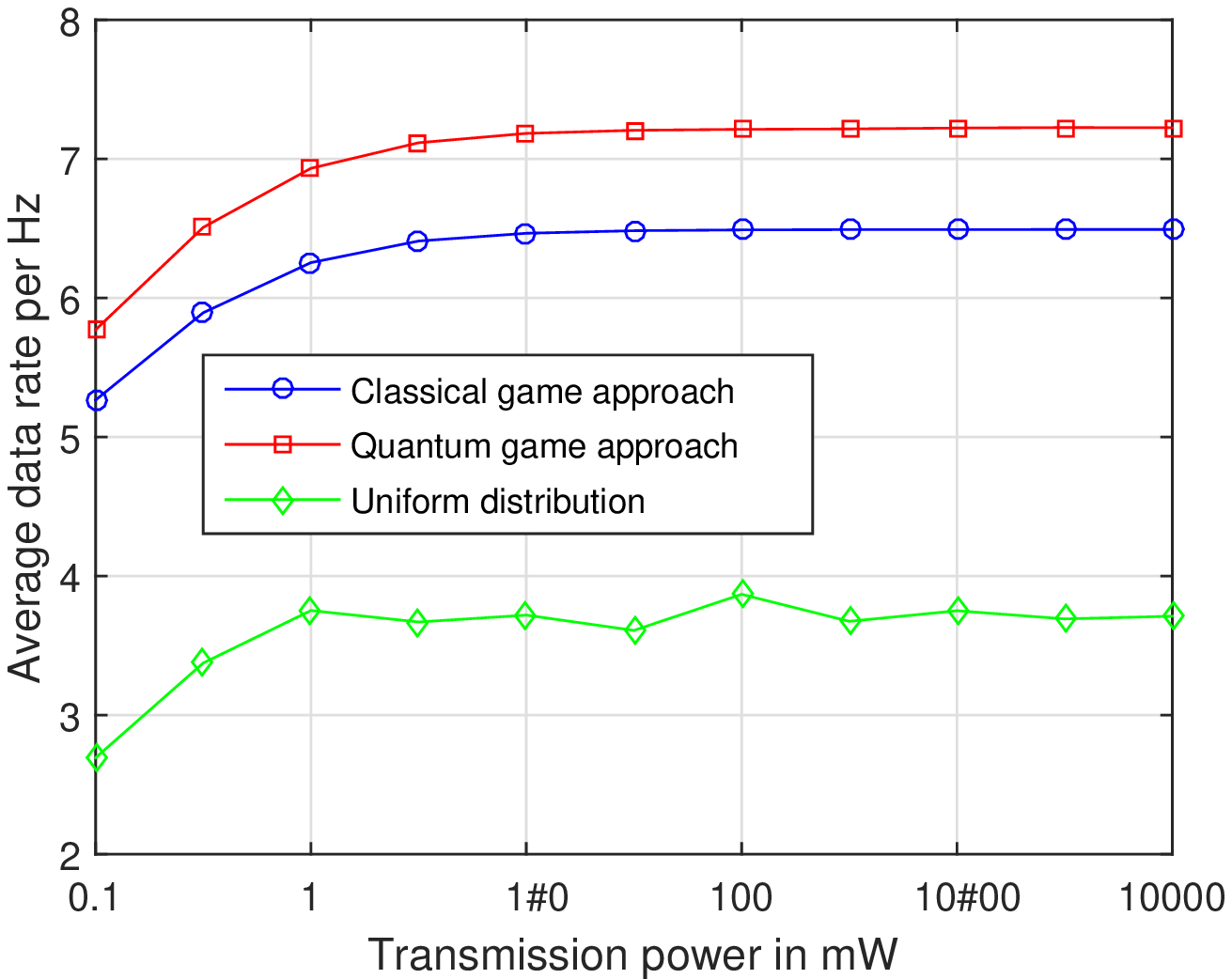}\vspace{-0.15cm}
\caption{\label{power}}
\end{subfigure}
\vspace{-0.25cm}
\caption{\small{\label{fig:both} Comparison of the average transmission rates resulting from the proposed conventional game, quantum game, and uniform approach
(a) as the degree of entanglement varies from $0$ to $\pi/2$, (b) when the QGT is guaranteed to improve the performance compared to standard game, as the degree of entanglement changes from $0$ to $\pi/2$, and (c) as the transmission power changes from $0.1$~mW to $10$~W.}
	}
\end{center}
\vspace{-0.6cm}
\end{figure*}


\vspace{-0.05cm}
\section{simulation results and analysis \label{sec5}}
For our simulations, a D2D wearable network over 60 GHz frequency band is deployed, and the transmission power for each device is $0.1$~Watts. The channel parameters are based on IEEE 802.11ad  \cite{standard}.
To clearly showcase the advantage of our proposed schemes, we compare them with a uniform beam alignment scheme in which the beamforming direction for each transmitter is randomly distributed in $[0,\pi/2)$.

In Fig.~\ref{entangle}, we show the average data rate resulting from all approaches as the parameter $\gamma$ varies. Here, recall that $\gamma$ measures the degree to which the quantum strategy spaces of all transmitters are entangled. Fig.~\ref{entangle} first shows that the classical game-theoretic approach yields significant improvements in the data rate, reaching up to $75\%$ gains, compared to the uniform case.
Moreover, by increasing $\gamma$, the average data rate that each transmitter can achieve using the quantum approach can be enhanced, while the transmission performance of the classical and the uniform beam alignment are unaffected.
Note that, when $\gamma$ equals to $0$, there is no entanglement between two transmitter's beam. Therefore, the strategies of two transmitters work independently, and no utility gain will be achieved compared with the non-quantum game. In contrast, as seen from Fig.~\ref{entangle}, in the maximally entangled case, where $\gamma = \pi/2$, the utility gain of the quantum beam alignment increases up to $10\%$ compared to the conventional game setting.

Note that the quantum game does not always bring a greater utility. When the NE of the classical approach is the social optimal, the quantum beam alignment cannot provide any rate improvement.
Since mathematically proving the conditions under which QGT is better can be difficult, in Fig.~\ref{only}, we show the average transmission rates resulting from \emph{only} the cases in which the quantum realm is guaranteed to have benefits. Thus, the results of Fig.~\ref{only} can be viewed as the maximal gains of the quantum game. Clearly, Fig.~\ref{only} shows that compared with the classical game, the proposed QGT approach can yield significant gains in terms of data rate, reaching up to $20\%$ (for $\gamma = \pi/2$), in the cases where quantum entanglement works.

Fig.~\ref{power} shows how the data rate varies for $\gamma = \pi/2$ and for a variable transmission power $P$. In this figure, first, we can see that the quantum game approach yields the best performance. Moreover, Fig.~\ref{power} also shows that, for a transmit power less than $10$~mW, the average data rate increases. However, when the transmit power exceeds $10$~mW, due to the increased interference, the overall data rate remains relatively constant at the NE, for both proposed approaches.

\begin{figure}[!t]
\begin{center}
	\vspace{-0.25cm}
\includegraphics[width=6.8cm]{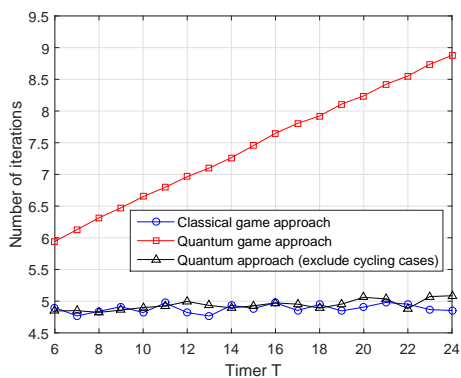}
\vspace{-0.1cm}
\caption{\label{converge}\small Average number of iterations needed for the convergence of the proposed approaches.}
\end{center}\vspace{-0.6cm}
\end{figure}

In Fig.~\ref{converge}, we evaluate the convergence of the proposed algorithms in terms of the number of iterations, as the timer $T$ varies. First, Fig.~\ref{converge} shows that, the average number of iterations of the classical game is about $5$, and does not increase with the timer.
In contrast, the average number of iterations of a general quantum game is higher and increases linearly with $T$.
From Fig.~\ref{converge}, we can see that the quantum game requires, on average, up to $9$ iterations, to converge. This increase is due to the occurrence of the nonconvergence in BR dynamics, where the iteration stops at the expiration of the timer.
Since the expansion of the strategy space in the quantum model increases the probability that BR dynamics does not converge, the average iterative times in the quantum approach grows with the value of $T$.
However, in Fig.~\ref{converge}, we can also see that, for the cases in which the QGT converges, the overall convergence time is similar to the classical game, despite the expanded strategy space. Therefore we can set $T = 5$ to de-emphasize the nonconvergence case. And the proposed approaches is shown to have reasonable convergence. \vspace{-0.15cm}

\section{Conclusion \label{sec6}}

In this paper, we have proposed a novel approach to analyze the optimal beam alignment of wearable mmW communications. The problem has been formulated as a noncooperative game between the transmitters of a wearable network. In this game, each transmitter chooses the optimal beamforming direction to maximize the data rate within its D2D communication pair.
To solve the proposed game, we presented an algorithm based on the best response dynamics, which enables the transmitters to reach a NE in a distributed way. To reap the gains in terms of the transmission rate, a quantum game model is proposed to expand the strategy space and allow transmitters to exploit quantum beamforming directions.
Simulation results show that the proposed game-theoretic approach significantly improve the data rate of up to $75\%$ compared to a uniform beam alignment strategy. Further, the QGT approach can yield up to $20\%$ of improvement, relative to the standard game.

\def\baselinestretch{0.8}
\bibliographystyle{IEEEtran}
\bibliography{references}

\end{document}